\documentclass[aps,prb, reprint, superscriptaddress, amsmath,amssymb,10pt]{revtex4-1}

\usepackage{graphicx,xcolor}
\usepackage{hyperref}
\usepackage{soul}

\newcommand{\IFNCNR}{Istituto di Fotonica e Nanotecnologie - Consiglio Nazionale delle Ricerche (IFN-CNR), p.za Leonardo da Vinci 32, 20133 Milano, Italy}
\newcommand{\POLIMI}{Dipartimento di Fisica - Politecnico di Milano, p.za Leonardo da Vinci 32, 20133 Milano, Italy}
\newcommand{\GAP}{Current affiliation: GAP-Quantum Technologies, Universit{\'e} de Gen{\`e}ve, Chemin de Pinchat 22, Switzerland}

\newcommand{\figref}[1]{Fig.~\ref{#1}}
\newcommand{\gradi}[0]{^{\circ}}

\makeatletter
	\renewcommand{\fnum@figure}{{\bf Figure \thefigure}}
\makeatother

\begin{document}

\title{Geometrically-controlled polarisation processing in an integrated photonic platform}

\author{Ioannis Pitsios}
\affiliation{\IFNCNR} \affiliation{\POLIMI}

\author{Farid Samara}
\thanks{\GAP}
\affiliation{\POLIMI}

\author{Giacomo Corrielli}
\affiliation{\IFNCNR} \affiliation{\POLIMI}

\author{Andrea Crespi}
\affiliation{\POLIMI} \affiliation{\IFNCNR}

\author{Roberto Osellame}
\email{roberto.osellame@polimi.it}
\affiliation{\IFNCNR}
\affiliation{\POLIMI}

\begin{abstract}
The polarisation of light is a powerful and widely used degree of freedom to encode information, both in classical and quantum applications. In particular, quantum information technologies based on photons are being revolutionised by the use of integrated photonic circuits. It is therefore very important to be able to manipulate the polarisation of photons in such circuits. We experimentally demonstrate the fabrication by femtosecond laser micromachining of components such as polarisation insensitive or polarising directional couplers, operating at 1550 nm wavelength, where the two opposite behaviours are achieved just by controlling the geometric layout of the photonic circuits, being the waveguides fabricated with the same irradiation recipe. We expect to employ this approach in complex integrated photonic devices, capable of a full control of the photons polarisation for quantum cryptography, quantum computation and quantum teleportation experiments.
\end{abstract}

\maketitle

\text{\bf Introduction}

Integrated optics is a very powerful platform to produce miniaturised complex photonic devices with improved scalability, robustness and suitability for field application.\cite{handbook} This approach greatly benefited classical optics communications, which is at the basis of today's information society. A similar trend is being adopted also for quantum optical devices, fostering an exponential increase in the layout complexity of integrated photonic circuits for various quantum applications, from computation to simulation.\cite{PolitiSilica,Silverstone,Marshall, Sansoni2010,szameitReview} Integrated quantum photonics is currently based on consolidated technologies, such as silicon on insulator \cite{Silverstone} and silica on silicon \cite{PolitiSilica}, as well as on more innovative approaches as waveguide writing by femtosecond laser micromachining (FLM) \cite{Marshall, Sansoni2010,szameitReview}. The two main advantages that FLM introduced in integrated quantum photonics are: the possibility to manipulate polarisation-encoded photons on-chip \cite{Sansoni2010} and unique 3D fabrication capabilities \cite{Crespi2016}. The former enables the direct on-chip transfer of many protocols already developed in quantum optics and based on polarisation encoding. The latter opens the way to innovative 3D layouts that can produce novel functionalities in more compact geometries.

The polarisation sensitivity of the FLM fabricated devices comes from the low ($ \sim 10^{-4} - 10^{-5}$) optical birefringence that is typically present in this type of waveguides. Depending on the material, birefringence in FLM typically originates from one or more of the following sources. It can be due to the creation of periodic nano-structures aligned orthogonally to the writing beam polarisation\cite{intrisic}, it can arise during fabrication due to asymmetric mechanical stresses\cite{stress}, or it may be due to ellipticity of the written waveguide cross section\cite{Keck,asymmetry}. While high birefringence is desirable to achieve effective polarisation manipulation in compact devices, on the other hand this same property induces strong delays between the two main polarisation components, thus rapidly destroying the original polarisation state where the information was encoded. The birefringence of the glass waveguides produced by FLM represents a very good compromise in this view, as it is low enough to produce an easily compensable effect on the polarisation state, even when propagating in several-centimetre-long waveguides, but it is high enough to produce a significant change in the coupling coefficients for different polarisations in a directional coupler (the integrated optics equivalent of a beam splitter). 

Taking advantage of this property, polarisation dependent components such as integrated polarising beam splitters \cite{Fernandes01} and polarisation retarders \cite{Fernandes02} have been presented for 1550 nm wavelength in fused silica. Polarisation waveplates \cite{Szameit01} have been demonstrated in the same material at 800 nm wavelength, while partially polarising beam splitters \cite{Crespi_ppl}, waveplates \cite{Corrielli01} and polarisation insensitive directional couplers \cite{Sansoni01} have been successfully demonstrated in alumino-borosilicate glasses at 785 nm wavelength. As it can be seen, scattered results in terms of material and wavelength are present in the literature. Here, we present a comprehensive study of polarisation sensitive and insensitive components produced by FLM in what are, in our opinion, very interesting conditions in terms of material and wavelength: the alumino-borosilicate glass EAGLE (Corning) \cite{NotaVetro} and operating wavelength at 1550 nm. In fact, the EAGLE glass easily allows achieving high-refractive-index waveguides at very high writing speeds, in the order of several cm/s, while the 1550 nm wavelength has very low losses in fiber transmission and is therefore extremely interesting in quantum optical communication \cite{Qcomm}, distributed quantum computing \cite{DQC} and quantum teleportation \cite{Qtel} experiments. In particular, we demonstrate a new approach to achieve polarisation insensitive directional couplers, we demonstrate high-performance polarising directional couplers and we show that also rotated waveplates can be produced in the same conditions. Finally, it should be noted that all these components are manufactured with the same fabrication recipe, this is very important in future complex photonic devices where it will be possible to combine polarisation sensitive and insensitive components on the same chip by only playing with their geometrical layout. 

\section*{Results}

\textbf{Directional couplers formalism} 
\begin{figure}[t]
\centering
\includegraphics[width=\linewidth]{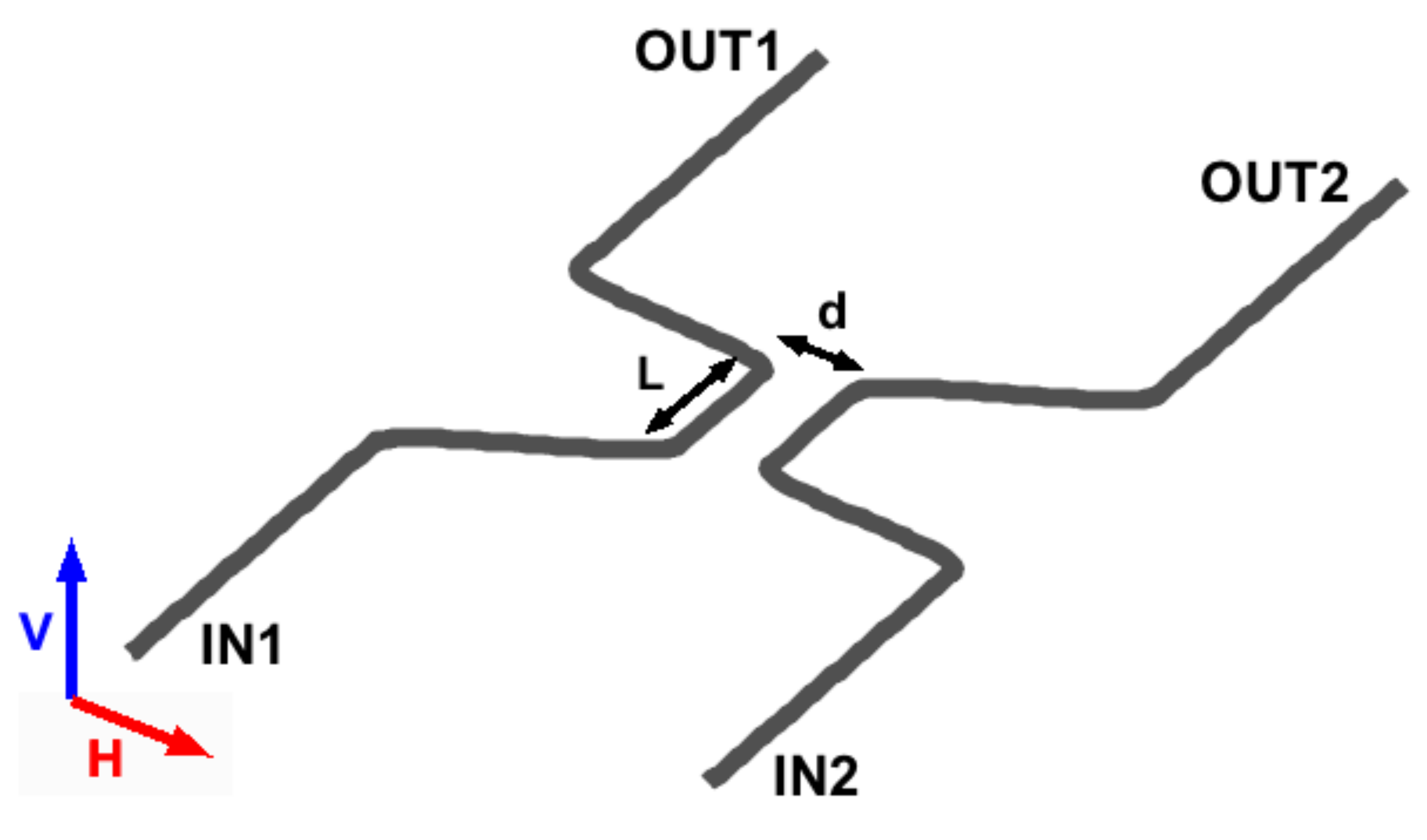}
\caption{Typical geometry of a directional coupler, lying on the horizontal plane. Two waveguides are brought close at a distance $d$ for a length $L$. Two input ports IN1 or IN2 and two output ports OUT1 and OUT2 can be identified. Orientations of horizontally (H) and vertically (V) polarisations with respect to the coupler plane are indicated.}
\label{f1}
\end{figure}
A directional coupler (DC) is the integrated-optics equivalent of the bulk-optics beam splitter. It consists of two waveguides that become proximal at a separation $d$ for a length $L$ (\figref{f1}). Within such a region, the two waveguides interact via evanescent-field coupling.\cite{Yariv,Huang} Light injected in either of the input ports (IN1 or IN2) thus partly remains in the same waveguide (BAR), while partly is transferred to the other arm (CROSS). In analogy with the bulk beam splitter we can define power transmission ($T$) and reflection ($R$) coefficients for a DC as:
\begin{align}
T &= \frac{P_\mathrm{CROSS}}{P_\mathrm{CROSS}+P_\mathrm{BAR}} &\qquad R &= \frac{P_\mathrm{BAR}}{P_\mathrm{CROSS}+P_\mathrm{BAR}}
\label{TR}
\end{align}
to be measured when light is injected in either of the input ports, and $P_\mathrm{BAR}$ and $P_\mathrm{CROSS}$ are the output powers on the two arms. 

If device losses are negligible or uniformly distributed along the waveguides, $T$ and $R$ do not depend on the choice of the input and oscillate as a function of the interaction length $L$ according to:\cite{Yariv}
\begin{align}
T &= \frac{\kappa^2}{\sigma^2} \sin^2 ( \sigma L + \phi_0), \notag\\
R &= 1 - T =  \frac{\Delta^2}{4 \sigma^2} + \frac{\kappa^2}{\sigma^2} \cos^2 ( \sigma L +\phi_0 ) 
\label{beatfull}
\end{align}
with the angular frequency $\sigma$ being defined as $\sigma^2 = \kappa^2 + \Delta^2/4$. In these expressions $\kappa$ is the coupling coefficient between the optical modes, $\Delta$ is their detuning in propagation constant,  and $\phi_0$ takes into account coupling occurring in the curved waveguide segments incoming and departing from the interaction region. 

Of course, if waveguides are identical $\Delta = 0$ and Eqs.~\eqref{beatfull} simplify to:
\begin{align}
T &=  \sin^2 ( \kappa L + \phi_0), \notag\\
R &=  \cos^2 ( \kappa L +\phi_0 ) 
\label{beat}
\end{align}

The coupling coefficient $\kappa$ depends on the overlap integral between the two waveguide modes, and thus it depends on separation $d$ between the two waveguides. For weakly guiding, femtosecond laser written waveguides, such dependency is well approximated by an exponential function\cite{Szameit2007} of the kind:
\begin{equation}
\kappa = \kappa_0 e^{-\frac{d}{d_0}} 
\label{coe}
\end{equation}
where $\kappa_0$ and $d_0$ are proper constants.
Evidently, by manipulating the interaction length and the separation of the waveguides, one can fully tune the coupler splitting ratio from null to unitary transmissivity or reflectivity.

{\bf Polarisation-splitting directional couplers}\\ 
A polarisation-splitting directional coupler (PDC) is the integrated equivalent of a bulk polarising beam splitter. The function of such a device is to separate orthogonal polarisations into different output arms. For instance, if vertically polarised light is coupled in port IN1 and all of it exits the device at OUT1, then horizontally polarised light injected in port IN1 should exit all from OUT2. As global figures of merit for the quality of a PDC we can define the logarithmic extinction ratios in transmission and reflection:
\begin{align}
\mathrm{ER}_T &= \left| 10 \log_{10} \frac{T_V}{T_H} \right| &\quad \mathrm{ER}_R &= \left| 10 \log_{10} \frac{R_V}{R_H} \right|
\label{extRat}
\end{align}
expressed in decibels. An ideal device would have infinite extinction ratios both in transmission and in reflection.

PDCs can be realized exploiting waveguide modal birefringence, as shown in the literature, which  reports partially polarising PDCs in borosilicate glass, working at 800~nm wavelength,\cite{Crespi_ppl} and fully polarising PDCs in fused silica, working at 1550~nm wavelength.\cite{Fernandes01} Here, as a first component of our polarisation manipulation platform, we address the optimization of PDCs in borosilicate glass working in the 1550~nm wavelength range.
\begin{figure*}[t]
\centering
\includegraphics[width=\textwidth]{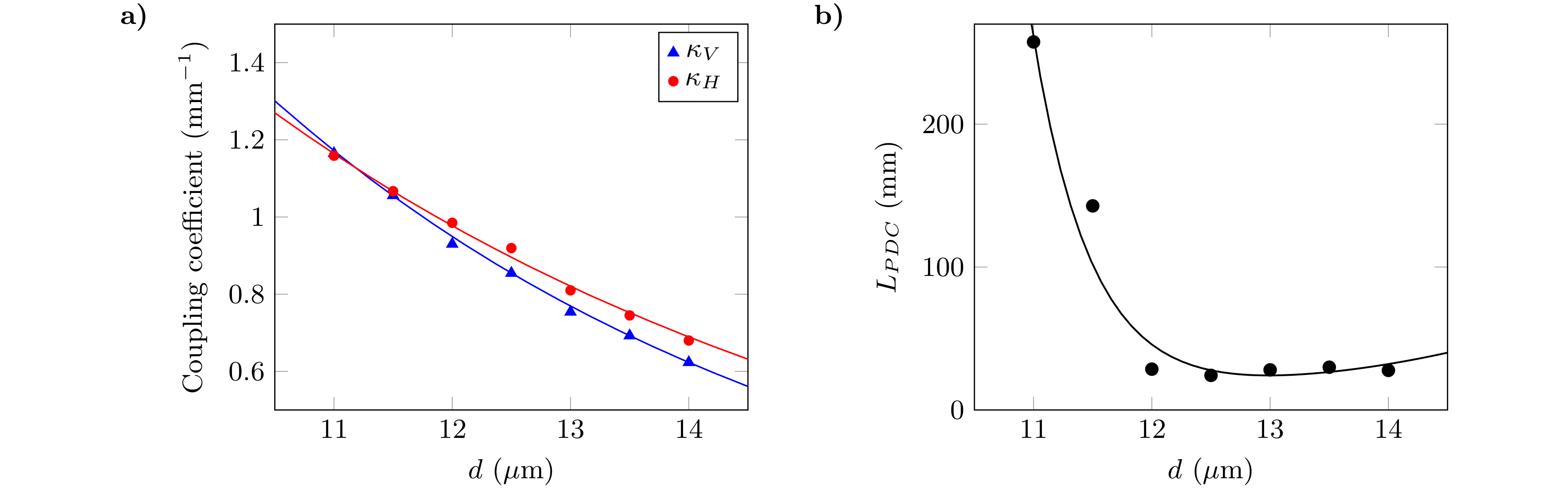}
\caption{(a) Experimentally measured coupling coefficients $\kappa_H$ (red circles) and $\kappa_V$ (blue triangles) for H and V polarised light respectively, as a function of the waveguide separation $d$ in the DCs. Solid lines are exponential fits. (b) Predicted interaction lengths $L$ for achieving a PDC, as a function of $d$, obtained by substituting in the second equation of \eqref{Lpbs} the experimental values of $\kappa_H$ and $\kappa_V$ (black circles) and their exponential fits (solid line).}
\label{f2}
\end{figure*}
\begin{figure*}[t]
\centering
\includegraphics[width=\textwidth]{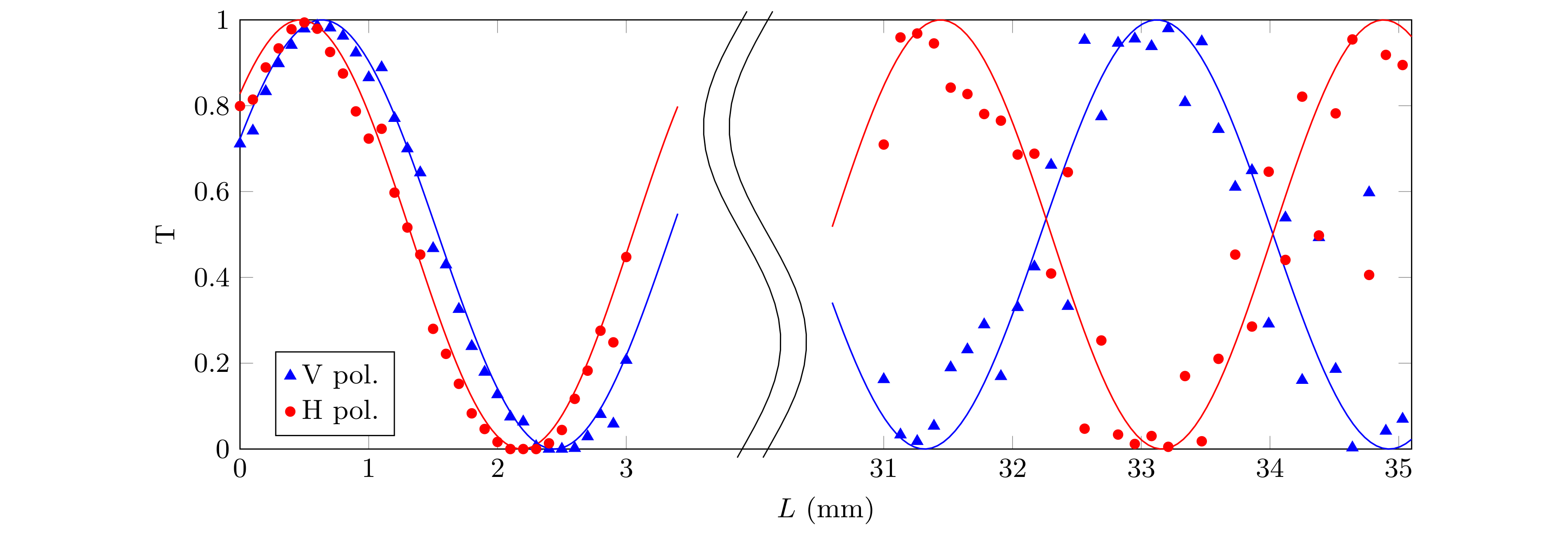}
\caption{Experimental measurements of the transmission $T$ of DCs with different interaction length $L$ and waveguide separation $d = 12.5 \mu$m, for H (red circles) and V (blue triangles) polarised input light. Solid lines are best-fitting  theoretical curves according to Eq.~\eqref{beat}.}
\label{f3}
\end{figure*}

The operation principle of a PDCs is simple. In general, because of waveguide birefringence, the power oscillation between the two waveguides in the DCs, described by \eqref{beat}, presents a different periodicity for different polarisation states (see the Methods for a more formal description of the polarisation behaviour of a generic two-port device). While for shorter interaction lengths $L$ the difference in splitting ratio may be minimal, the polarisation dependence may become larger with increasing $L$ where the two beatings accumulate larger phase difference. In particular, for a proper choice of the DC geometrical and coupling parameters, one may reach a point in which the two oscillations are in anti-phase, which is indeed the condition for a PDC. From equations \eqref{beat}, and assuming for simplicity $\phi=0$, we can write mathematically the PDC condition as follows:
\begin{equation}
\left \lbrace
\begin{array}{l}
\kappa_V L = m \pi \quad \vee \quad \kappa_V L = \left(m + \dfrac{1}{2}\right) \pi \\
\kappa_H L = \kappa_V L \pm \dfrac{\pi}{2}
\end{array}
\right.
\label{PDCcondition}
\end{equation}
where $m$ is integer and we have differentiated $\kappa_V$ and $\kappa_H$ as coupling coefficients for two orthogonal polarisations, here considered vertical (V) and horizontal (H). The choice of the first equation (with or without the $\frac{1}{2} \pi$ term) depends on the desired behaviour for the V~polarisation, namely if one wants it totally reflected or totally transmitted. The sign in the second equation will be plus (minus) if $\kappa_V$ is smaller (larger) than $\kappa_H$. It is not difficult to observe that satisfying exactly these equations requires $\kappa_H/\kappa_V$ to be a ratio of integer numbers. Namely, \eqref{PDCcondition} can be elaborated as follows:
\begin{equation}
\left \lbrace
\begin{array}{l}
\dfrac{\kappa_H}{\kappa_V} = \dfrac{2 m + 1}{2 m} \quad \vee \quad \dfrac{\kappa_H}{\kappa_V} = \dfrac{2 m + 1}{2 m + 2}\\[15pt]
L = \dfrac{\pi}{2\left(\kappa_H - \kappa_V\right)}
\end{array}
\right.
\label{Lpbs}
\end{equation}
where we have assumed $\kappa_H > \kappa_V$.
In principle the ratio $\kappa_V/\kappa_H$ could be modulated continuously by the choice of the interaction distance $d$, and it should be possible to fulfill \eqref{Lpbs} up to an arbitrary high precision by a careful optimization of both $d$ and $L$. In practice, the non-perfect reproducibility of the fabrication technology would make this process uneffective.
Therefore, as it will be discussed later in detail, we will choose a value for $d$ sufficiently close to the optimum, and then we will proceed in optimizing $L$, concentrating in particular on one of the two extinction ratios (either $\mathrm{ER}_T$ or $\mathrm{ER}_R$), thus sligthly compromising on the other one. 

\begin{figure}[t]
\centering
\includegraphics[width=\linewidth]{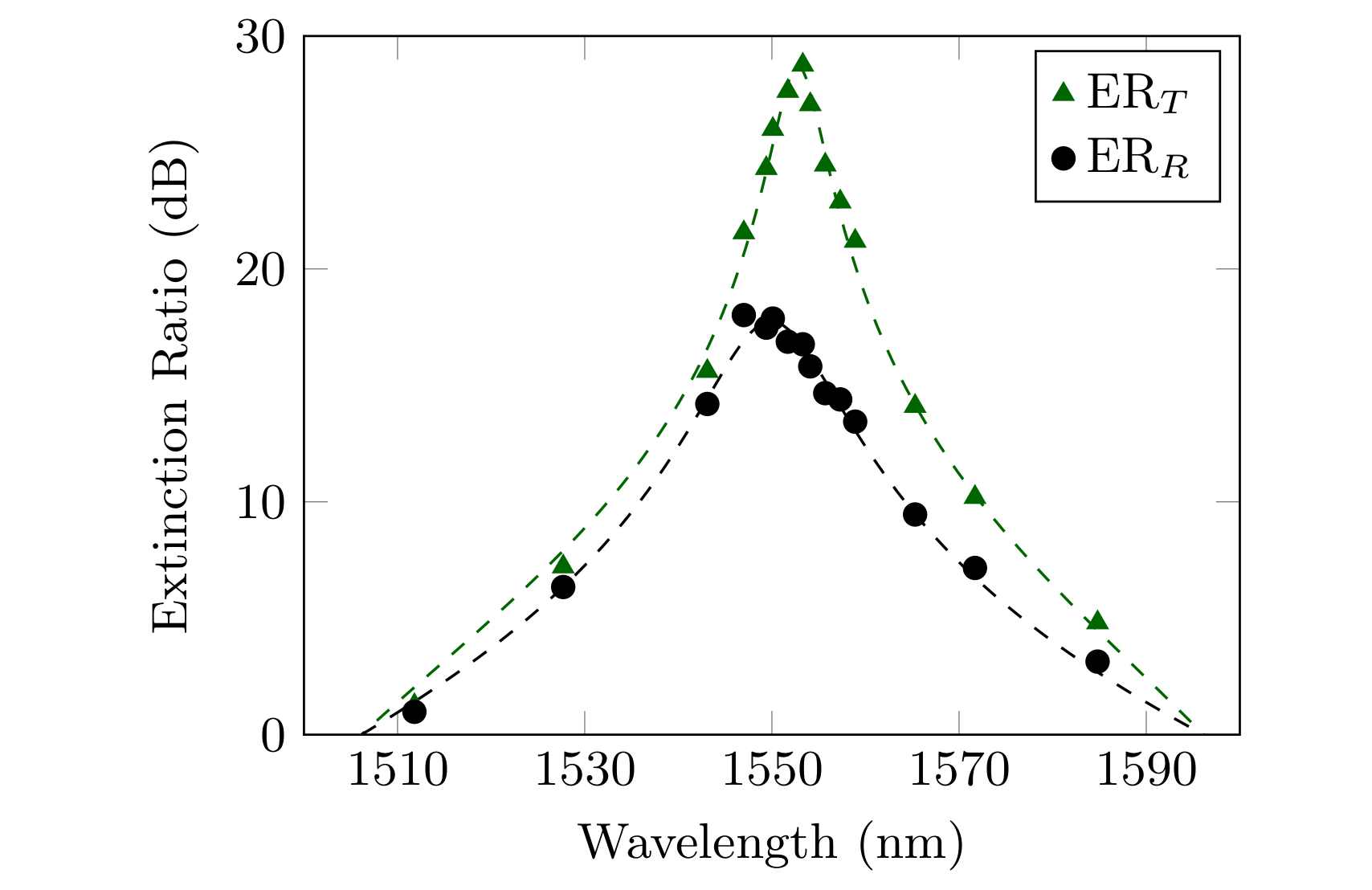}
\caption{Spectral characterisation of the transmission and reflection extinction ratios for a PDC fabricated with geometrical parameters $d$~=~12.5~$\mu$m and $L = 33.2$~mm. Dashed lines are best-fitting curves from an analytic model in which we assumed a linear dependency of the coupling coefficients $\kappa_{H}$ and $\kappa_{V}$ from the wavelength and $T_V$ and $T_H$ can oscillate with a generic non-unitary visibility.} 
\label{f4}
\end{figure}
Our femtosecond laser written waveguides in borosilicate glass, supporting a single mode at 1550~nm wavelength, yield quite a low modal birefringence $b = 3.6 \cdot 10^{-5}$ (see the Methods section for details on the fabrication and characterisation process). The birefringence axis is oriented vertically, due to the symmetry constraints of the waveguide writing process.\cite{Corrielli01} This gives distinct guided modes for H and V polarisations, with a slightly different size and shape. The measured H mode size (1/$e^2$) is 15.6~$\times$~15.1~$\mu$m$^2$ while the V mode size is 15.7~$\times$~15.5~$\mu$m$^2$. Such mode-size difference results in different $\kappa_0$ and $d_0$ values in \eqref{coe} depending on the light polarisation, which produce in general a polarisation dependent $\kappa$ value for a given interaction distance $d$ of a directional coupler. As a first step to the realisation of a PDC, we characterised thoroughly the dependence of the coupling coefficients $\kappa_H$ and $\kappa_V$ on the interaction distance $d$. To this purpose, we fabricated DCs with the same length and separations varying from 11~$\mu$m to 14~$\mu$m. The output power distribution for each polarisation was collected and the different coupling coefficients were retrieved by inverting relations \eqref{beat}. Results are reported in \figref{f2}a.

An estimate of the interaction length $L_{PDC}$ required to achieve a PDC can be given by simply substituting the measured values of $\kappa_H$ and $\kappa_V$, as a function of $d$, in the second equation of the condition \eqref{Lpbs} (see \figref{f2}b). It is important to note that $L_{PDC}$ becomes large both when $d$ is too short and the coupling coefficients for the two polarisations tend to be similar, and when $d$ is too large and both coupling coefficients become small (even though they relatively differentiate). Thus, there is an optimum value for $d$ which gives the shortest device. According to the experimental data and their best fit, in our case such value should be around $d_{opt} = 12.5$~$\mu$m, corresponding to a $L_{PDC} \sim 30$~mm. 

Once the optimum interaction distance $d_{opt}$ was chosen, in order to measure more accurately $\kappa_H$, $\kappa_V$, $\phi_H$ and $\phi_V$ for that distance, and to have a better estimate of the actual $L_{PDC}$, we fabricated 31 DCs spanning $L = 0 \div 3$~mm. The left part of the graph in \figref{f3} reports the measured transmission $T$ for these devices, for the two polarisations as a function of $L$. By extrapolating the sinusoidal trends to longer lengths we identified $L_{PDC} \simeq 33$~mm, in agreement with the previous estimate. Subsequently, we fabricated 40 devices to scan $L = 31 \div 35$~mm. Measured transmission values for these devices are reported in the right part of the graph in \figref{f3}. 

The device showing the best performance was the one with $L = 33.2$~mm, which yielded transmissions $T_V=0.980$ and $T_H=0.002$ for the two polarisations respectively, corresponding to $\mathrm{ER}_T = 26.9$~dB and $\mathrm{ER}_R = 17$~dB at the design wavelength of 1550~nm; this device was chosen for an in-depth characterisation. Figure~\ref{f4} reports the spectral characterisation of the extinction ratios. The highest extinction for the transmission reached the value of $\mathrm{ER}_T = 28.8$~dB at 1553.3~nm wavelength while for the reflection the highest value was $\mathrm{ER}_R = 18.9$~dB at 1549.4~nm. Extinction ratios greater than 15~dB were indeed observed in transmission for a bandwidth of about 14.2~nm. The slight difference in the wavelengths for the maximum values of transmission and reflection extinction ratios can be attributed to the approximated optimisation procedure discussed above, which focuses mainly on one polarisation. Higher values for the extinction ratios might be found with a finer tuning of the interaction length $L$. However, a limitation in the achievable value of the extinction ratio could be given by waveguide non-uniformities, or by slight imbalances of the optical properties between the two waveguides. These may arise e.g. from fluctuations in the power of the waveguide writing laser, or by stress induced on the first waveguide by the writing process of the second one, as they are inscribed sequentially. 
In particular, even a small, uniform detuning between the propagation constants of the two waveguides forbids\cite{Yariv} to reach exactly null reflection values; this may explain the fact that the maximum observed values for $\mathrm{ER}_R$ is lower than the maximum observed $\mathrm{ER}_T$.

{\bf Polarisation-insensitive directional couplers}
A polarisation-insensitive coupler (PIC) is a device able to split orthogonal polarisations equally. Insensitivity to polarisation is often fundamental in applications that involve polarisation encoding of quantum states, or polarisation-entangled photons, because a polarisation sensitive behaviour would definitely introduce some kind of distinguishability and thus undermine the quality and purity of the quantum states. PICs have been demonstrated recently\cite{Sansoni01} with femtosecond laser written waveguides in borosilicate glass substrate, operating in the 800~nm wavelength range. While the component waveguides retained the same optical properties and birefringence of those used elsewhere\cite{Crespi_ppl} for demonstrating partially polarising couplers, there the use of a peculiar three-dimensional geometry enabled an equalisation of the coupling coefficients for the two polarisations, and hence the achievement of polarisation insensitivity. The operation principle of those devices is briefly described as follows. For weakly coupled modes, the coupling coefficients $\kappa_{V,H}$ are directly proportional to the overlap integrals of the two H- and V-polarised waveguide modes. Since the H- and V-polarised modes present a different ellipticity, if two coupled waveguides lie on a plane tilted with a proper angle with respect to the horizontal, it may be possible to reach a condition for which the two overlap integrals become equal and thus $\kappa_H = \kappa_V$.

\begin{figure}[t]
\centering
\includegraphics[width=\linewidth]{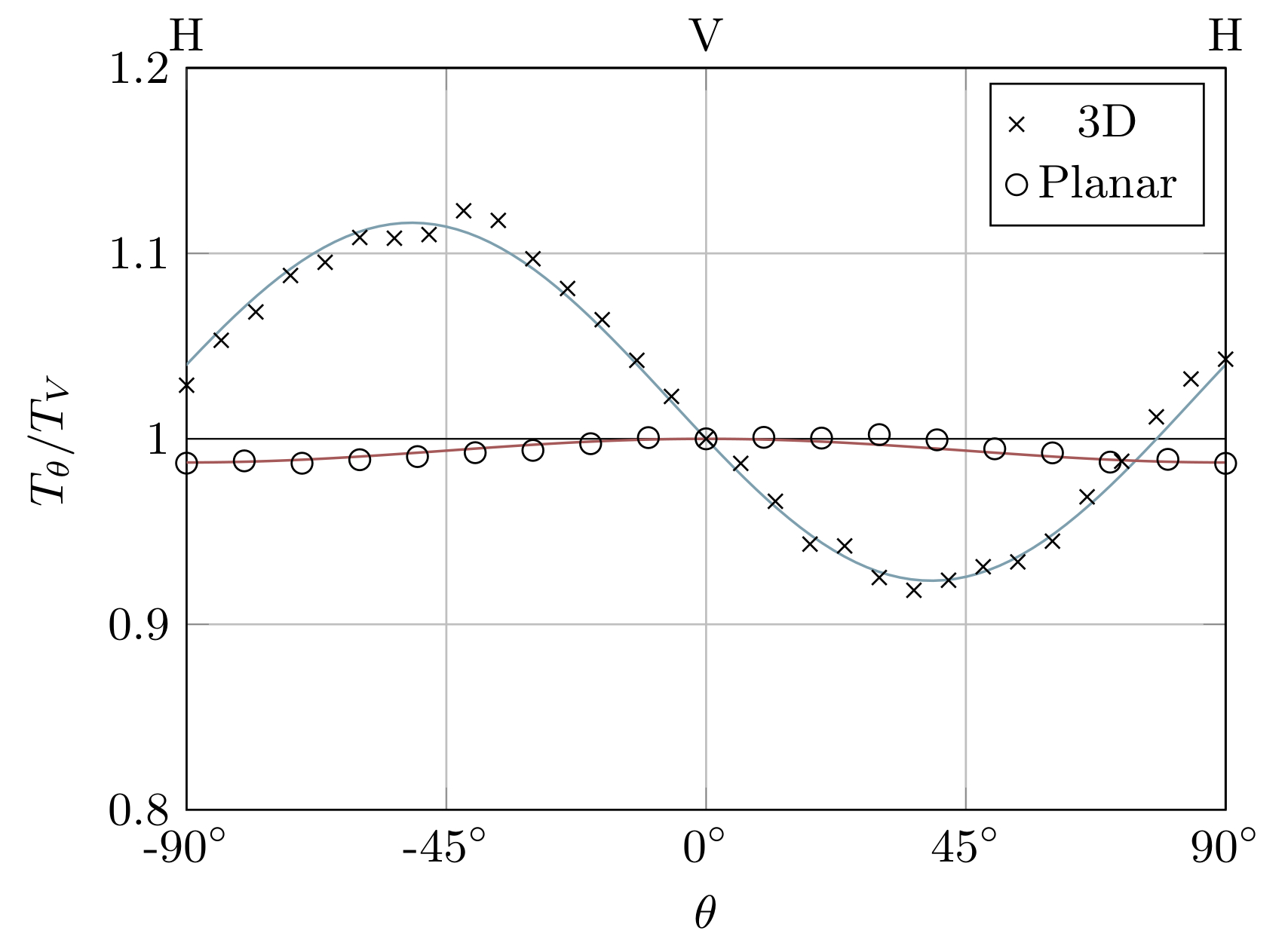}
\caption{Ratio between the transmission $T_\theta$ of a DC, measured by injecting linearly polarised light at a polarisation angle $\theta$ with respect to the vertical one, and the transmission $T_V \equiv T_{0^\circ}$, measured for vertical polarisation. Typical results are reported both for a three-dimensional coupler (crosses), realized with the method of Ref.~\onlinecite{Sansoni01}, and a planar one (circles), realised with the method discussed in the text. Solid lines are best fits according to Eqs.~\eqref{genericT} and \eqref{Tbcoupler}, discussed in the Methods section.}
\label{f6}
\end{figure}

\begin{figure*}[t]
\centering
\includegraphics[width=\textwidth]{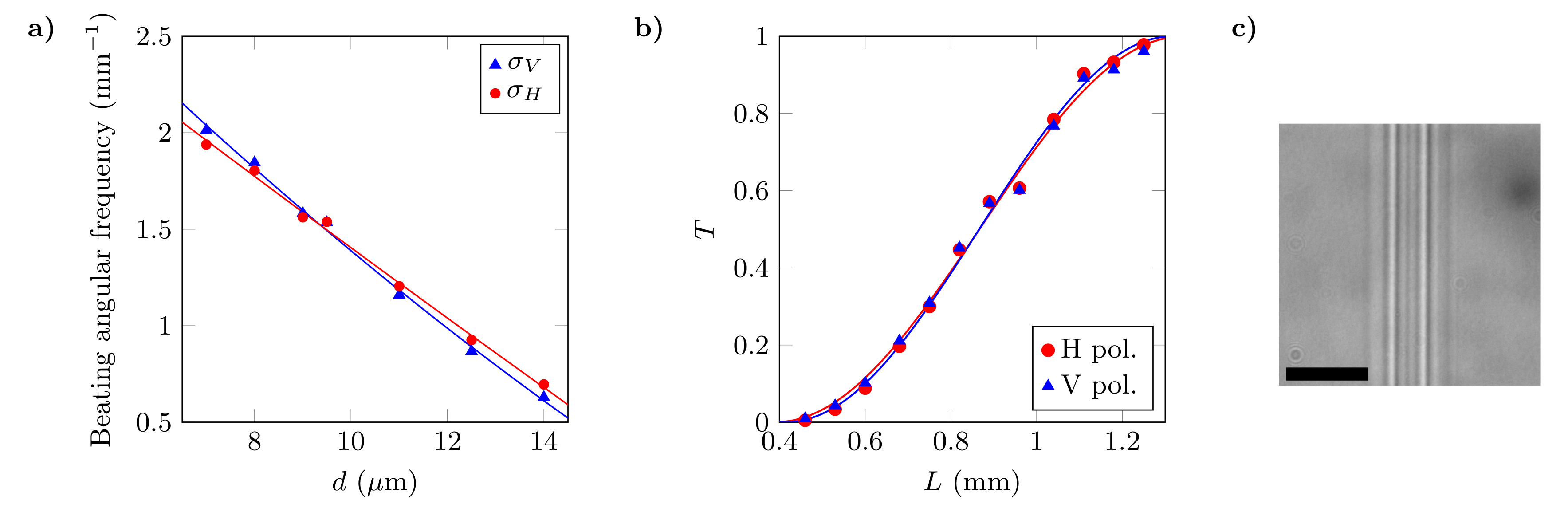}
\caption{(a)  Experimentally retrieved values for $\sigma_H$ (red circles) and $\sigma_V$ (blue triangles), for inter-waveguide separations comprised between 7~$\mu$m and 11~$\mu$m. Solid lines are second-order polynomial fits, since for such short interaction distances the exponential approximation \eqref{coe} is not accurate. (b) Transmission $T$ for PICs built with $d$~=~8~$\mu$m and $L = 0.4 \div 1.3$~mm, for H- (red circles) and V- (blue triangles) polarised input light. Sinusoidal best fits according to \eqref{beat} are also plotted. (c) Microscope image of the interaction region of one of the PICs with $d$~=~8~$\mu$m: the two laser-written waveguides have central regions with higher index contrast (whiter in the picture) which are still well apart, but more outer regions practically touch. The scale-bar is 20~$\mu$m.}
\label{f5}
\end{figure*}

At first, we tried to follow the above approach to devise PICs working at 1550~nm. Indeed, the waveguides used in the present work differ from those reported in Refs.~\onlinecite{Crespi_ppl,Sansoni01} only for the higher inscription power (because of the larger operation wavelength, the waveguide size has also to be larger). To this purpose, we fabricated and characterised several DCs with different interaction distances $d$ and tilting angles of the coupler plane, observing for some of them the desired equality of splitting ratio for both H- and V-polarised input light. However, a more careful study of their polarisation performance showed the behaviour reported in \figref{f6}: light injected with H or V polarisation direction produces an equal splitting ratio, but the splitting ratio changes when the light injected in the coupler is linearly polarised along a more generic angle. Additional measurements also showed that H- or V-polarised input light underwent a slight polarisation rotation while propagating in the couplers. These phenomena can be explained only by assuming that the birefringence axis of the waveguides does not remain unaltered and vertically oriented in the whole directional coupler, but it changes orientation in at least some part of the device (see the Methods section on the formalism). From a physical point of view, the rotation of the birefringence axis in the coupler can be explained by a mechanical stress induced on the first waveguide, within the coupling region, by the inscription of the second waveguide. In fact, Heilmann et al.\cite{Szameit01} recently observed similar rotation of the waveguide birefringence axis when inscribing close-by traces at an angle. We observed the same effect in an experiment where an integrated optical circuit fabricated by FLM was used for the spatial multiplexing of orthogonal polarisation light states, for quantum cryptography applications\cite{Vest2015}. Here, very compact 50-50 PICs ($d$~=~7$\mu$m) for 850 nm light were inscribed in borosilicate glass using the out-of-plane geometry, which caused small polarisation distortion at the circuit output. Due to the highly demanding tolerances on the output polarisation state required by the specific experiment, a compensation of this non-ideal behaviour was implemented. On the contrary, we never observed significant effects related to this phenomenon in the experiments of Refs.~\onlinecite{Crespi_ppl,Sansoni01}, as the tiny polarisation rotation induced by the couplers does not alter appreciably the circuit output power distribution.

An alternative route for designing PICs can be individuated by a more careful inspection of \figref{f2}a. In fact, one can notice that for shorter and shorter interaction distances $d$ the coupling coefficients for the two polarisations tend to become similar. If, by further decreasing $d$, the two coupling coefficients not only become more similar, but also exchange magnitudes, there will be a point in which they cross. Namely, in that point the two coefficients $\kappa_H$ and $\kappa_V$ will be equal. 

In order to study the validity of this hypothesis, we fabricated and characterised sets of directional couplers with waveguide separations $d$ between 7~$\mu$m and 11~$\mu$m, and lengths $L$ spanning from 0~mm to 1.5~mm. These couplers keep the planar geometry of \figref{f1} and thus differ from the PDCs described previously only for the different geometrical parameters. For each group of devices we retrieved, for the two polarisations, the beating periodicity for the transmission and reflection of the coupler as a function of the length $L$; namely, we could retrieve the quantity $\sigma$ in Eqs.~\eqref{beatfull}. We prefer here to describe the power beatings in terms of $\sigma$ and not in terms of $\kappa$ because we cannot exclude the presence of a small detuning between the propagation constants of the two waveguides, as we will discuss below. We observe in \figref{f5}a that indeed $\sigma_V = \sigma_H$ for $d \simeq$~9~$\mu$m, while $\sigma_V > \sigma_H$ for shorter distances and vice versa for larger ones.

In practice, to realise a working PIC it is not sufficient to set $d$ as close as possible to the distance at which $\sigma_H = \sigma_V$. In fact, in the actual device coupling happens both in the central region and in the approaching curved waveguide segments. The overall behaviour results from different coupling contributions occurring at different separations: the central region will likely have to be built with a short distance $d$ for which $\sigma_V > \sigma_H$, to balance the opposite tendency of the coupling occurring in the curved parts where $\sigma_V < \sigma_H$.
Figure~\ref{f5}b shows the measured transmission values for DCs with $d = 8 \mu$m and $L = 0.4 \div 1.3$~mm. The absolute transmission differences for the two polarisations reaches impressively low values down to about $10^{-4}$, for the device length $L$ that most balances the two different coupling contributions.

We analysed the performance of these devices for different linear polarisation states at the input. As shown in \figref{f6}, the transmission of these DCs remains practically constant for every polarisation angle.  We also checked that no polarisation rotation occurs when injecting H- or V-polarised light, thus indicating that the birefringence axis of the waveguides maintains the same vertical orientation throughout the whole device. The small oscillation with $\theta$ still discernible in \figref{f6} can be explained by a residual difference in the coupling coefficients of the two polarisations but is consistent with a birefringence axis that remains fixed and vertically oriented (see the Methods section). It is finally important to note that, notwithstanding the very tight separation between the waveguides in the interaction region (see \figref{f5}c), we didn't observe any degradation of the waveguide quality or any additional loss in these DCs with respect to analogous devices with the same waveguide length but larger coupling separation.

The physical reason for such a polarisation insensitivity is likely to be an alteration of the optical properties of the first inscribed waveguide of the coupler, caused by the inscription of the second. In particular, the mechanical stress caused by the second inscribed trace may affect the modal birefringence of the pre-existent waveguide, making it decrease and possibly making it change its sign, for closer and closer waveguides. This effect would be analogous (but opposite in sign) to that reported in Ref.~\onlinecite{Fernandes03}, where the authors observe a change in the birefringence magnitude when other modification traces are inscribed next to the waveguides, in fused silica substrate. On the contrary, the birefringence of the second inscribed structure would likely remain the same one of the isolated waveguide, since the laser induced material-modification process involves high temperatures that should anneal any pre-existent mechanical stress in the region.

Such alterations of the first waveguide cause a change in its polarisation mode sizes (if considered as an isolated structure) and thus directly influence the coupling coefficients $\kappa_H$ and $\kappa_V$. In addition, they may affect the relative detuning between the waveguides, $\Delta_H$ and $\Delta_V$ for the two polarisations. In any case, results in \figref{f5} show that, for a given distance $d$, it is possible to achieve the condition:
\begin{equation}
\kappa^2_H + \Delta^2_H / 4 = \kappa^2_V + \Delta^2_V / 4
\end{equation}
namely, an equalisation of the beating frequencies $\sigma_H$ and $\sigma_V$.
Since transmission values close to unity are observed experimentally, a possible detuning $\Delta$ should actually be much smaller than the coupling coefficient $\kappa$ (see Eqs.~\eqref{beatfull}). This can be reasonable given the very short waveguide separation and high coupling coefficients involved.

\section*{Discussion}
We have demonstrated fully polarising DCs, fabricated by femtosecond waveguide writing in borosilicate glass, operating in the telecom wavelength range. These devices present extinction ratios between the two polarisations higher than 25 dB at the design wavelength of 1550~nm, maintaining extinction ratios higher than 15~dB in a range of $\sim 14$~nm.
Additionally, we developed a new architecture that allows to generate highly polarisation insensitive couplers with a planar geometry,using waveguides with the same irradiation parameters as the ones used for the polarising devices.
\begin{figure}[t]
\centering
\includegraphics[width=\linewidth]{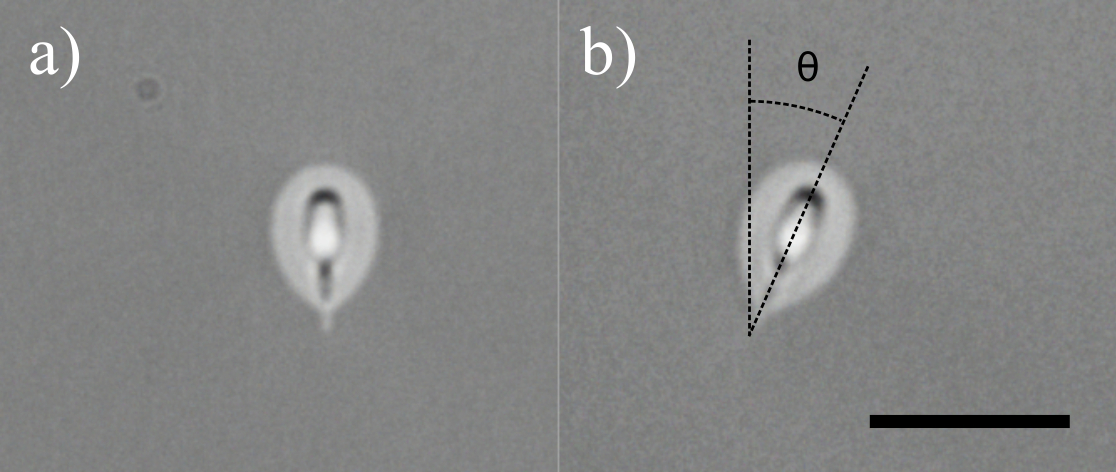}
\caption{Cross section of single mode optical waveguides at 1550 nm wavelength, written in borosilicate glass with the technique presented in Ref.~\onlinecite{Corrielli01}. Waveguide a) is fabricated with vertical optical axis while waveguide b) is fabricated with the optical axis tilted by $\theta$~=~25.6$\gradi$. The measured value of birefringence for the two waveguides is $b_a$~=~2.8$\cdot 10^{-5}$ and $b_b$~=~3.4$\cdot 10^{-5}$. Scale-bar is 20 $\mu$m.}
\label{f7}
\end{figure}
Importantly, this means that it is possible to combine in a single circuit both polarisation dependent and polarisation independent devices operating at the telecom wavelength, by just tuning the geometrical parameters of the components. 
These results become even more significant, as we have proven experimentally that the same technique presented in Ref.~\onlinecite{Corrielli01} for writing waveguides with rotated optical axis can be straightforwardly extended to telecom C-band devices (see figure \ref{f7}). This permits to implement arbitrary polarisation manipulation on-chip, and completes the set of functionalities required for the development of a polarisation-based integrated architecture operating at 1550~nm.

Improved performances and compactness of the devices might be pursued by varying the irradiation parameters used for waveguide writing. Thermal annealing techniques\cite{Arriola} may also be investigated to ameliorate waveguide uniformity, enhance the achievable extinction ratios of PDCs, and at the same time decrease both propagation losses and coupling efficiency with fibers.

These results extend in the telecom wavelength band the capability of femtosecond laser written circuits to handle and manipulate the polarisation degree of freedom, which was reported in the recent years in the 800~nm wavelength region.\cite{Sansoni2010, Crespi_ppl, Sansoni01} This paves the way to the integrated manipulation of polarisation encoded photonic qubits and polarisation entanglement at this interesting wavelength range. Notably, and at difference to other integrated platform such as silicon photonic circuits, femtosecond laser written waveguides present very good mode-matching and easy interfacing with standard optical fibers. These possibilities could thus promote the application of integrated-optics technology, with all its advantages in therms of stability and compactness, to practical demonstrations of quantum communication protocols within standard fiber networks. In addition, the control of polarisation, combined with the phase stability intrinsic to waveguide circuits, may open further perspectives to integrated quantum optics experiments with hyper-entangled photons.\cite{Ciampini2016}

\section*{Methods}

{\bf Device fabrication and characterisation}
The devices were fabricated using a femtosecond Yb:KYW cavity-dumped mode-locked oscillator, emitting pulses of 300~fs at 1~MHz repetition rate, with a wavelength $\lambda$~=~1030nm. The laser beam was focused 170~$\mu$m beneath the surface of Corning Eagle2000 alumino-borosilicate glass, using a 50$\times$ microscope objective of 0.6 NA. The translation of the sample was performed by computer controlled  air-bearing stages (Aerotech FiberGLIDE 3D). Irradiation parameters for the inscription of single-mode waveguide for 1550~nm operation wavelength were 370~nJ pulse energy and 40~mm~s$^{-1}$ translation speed. Measured propagation losses are 0.3 dB~cm$^{-1}$. The radius of curvature employed for the segments of the DCs was 90~mm, giving additional bending losses of 0.4 dB~cm$^{-1}$. Measured coupling losses to standard single mode fibers at 1550~nm are 0.4 dB. To measure waveguide birefringence we adopted the method described in the Supplemental Material of Ref.~\onlinecite{Sansoni2010}: namely, we injected light with different polarisation states in the waveguides and we characterised the polarisation states at the output, then best fitting the birefringence value that provided the observed transformation.

{\bf Generic polarisation description of two-waveguide devices}
A generic linear integrated-optics device having two input waveguides and two output waveguides, where two polarisation modes are considered for each distinct input/output port, operates as a linear transformation from four input optical modes to four output optical modes. In the lossless case, this transformation is unitary. If the waveguides at the input and output are birefringent with a vertical axis, the two polarisation modes can be assumed as linearly polarised modes, with H or V orientation. 

In detail, the operation of the device can be described as a unitary matrix:
\begin{align}
U = \left[
\begin{matrix}
a_1 & a_2 & b_1 & b_2\\
a_3 & a_4 & b_3 & b_4\\
c_1 & c_2 & d_1 & d_2\\
c_3 & c_4 & d_3 & d_4
\end{matrix}
\right]
\label{uMatrix}
\end{align}
applied to the vector of the mode field amplitudes:
\begin{align} \bar{E} = 
\left[
\begin{matrix}
H_1\\
H_2\\
V_1\\
V_2
\end{matrix}
\right]
\label{modesVector}
\end{align}
where $H_i$ and $V_i$ are the modes of the port $i$, respectively H- or V- polarised. In a quantum description, the same formalism can be used if the elements of the vector \eqref{modesVector} are considered as the mode annihilation operators.

It can be noted the terms $b_j$ and $c_j$ in the matrix \eqref{uMatrix} makes the H- and V- polarised modes interact and exchange optical power. All those terms should then be vanishing if the birefringence axis of the waveguides remain uniform and vertically oriented in the whole device: in fact the H and V polarised modes remain separated and never exchange power. In that case, the device behaves as two independent couplers operating on the H and V polarisation respectively:
\begin{align}
U_c = \left[
\begin{matrix}
r_H & \iota \, t_H & 0 & 0\\
\iota \, t_H^* & r_H^* & 0 & 0\\
0 & 0 & r_V e^{\iota \beta} & \iota \, t_V e^{\iota \beta}\\
0 & 0 & \iota \, t_V^* e^{\iota \beta} & r_V^* e^{\iota \beta}
\end{matrix}
\right]
\label{cMatrix}
\end{align}
where $r_H$, $r_V$, $t_V$ and $t_H$ are complex coefficients, $\beta$ is a phase term which accounts for birefringence effects (which delay one polarisation more than the other after a propagation through the whole device), and global phase terms on the whole matrix are neglected. If such device physically correspond to a DC as the one in \figref{f1}, built of two identical waveguides, one can write more precisely $r_H = \cos ( \kappa_H L + \phi_H)$, $t_H = \sin ( \kappa_H L + \phi_H)$, $r_V = \cos ( \kappa_V L + \phi_V)$, $t_V = \sin ( \kappa_V L + \phi_V)$, using notation consistently with the one used in the text.

{\bf Coupler operation for generic linearly polarised input light}
Linearly polarised coherent light at a generic orientation $\theta$ with respect to the V axis, injected in one input port of the coupler (e.g., input port 1), is described by a vector written as:
\begin{equation}
\bar{E} = E_0 \cdot
\left[
\begin{matrix}
\sin \theta\\
0\\
\cos \theta\\
0
\end{matrix}
\right]
\end{equation}
where $E_0$ is the input field amplitude and the formalism of Eqs.~\eqref{uMatrix},\eqref{modesVector} is  adopted.

It is not difficult to show that, in the case of generic device described by a matrix \eqref{uMatrix}, the power transmission at the output 2, defined as in \eqref{TR}, takes the form:
\begin{equation}
T = C_1 + C_2 \cos 2 \theta + C_3 \sin 2 \theta
\label{genericT}
\end{equation}
where $C_1 = \frac{1}{2} \left( |a_3|^2+|b_3|^2+|c_3|^2+|d_3|^2 \right)$, $C_2 = \frac{1}{2} \left( |a_3|^2+|b_3|^2-|c_3|^2-|d_3|^2 \right)$ and $C_3 = \mathfrak{R}\lbrace a_3 b_3^* + c_3 d_3^* \rbrace$.
Namely, the power transmission oscillates generically with the polarisation orientation $\theta$, with a periodicity of 180$^\circ$.

On the other hand, if we consider a coupler with fixed vertical birefringence axis, whose matrix takes the form \eqref{cMatrix}, the coefficient $C_3$ in \eqref{genericT} vanishes, and $T$ becomes:
\begin{equation}
T = \frac{1}{2} \left[ |t_H|^2 + |t_V|^2 + \left( |t_V|^2 - |t_H|^2 \right) \cos 2 \theta \right].
\label{Tbcoupler}
\end{equation}
Actually, $T$ oscillates sinusoidally between the transmission values observed for pure H and V polarisations, having the maxima or minima exactly in $\theta = 0^\circ$ and $\theta=\pm 90^\circ$.

\textbf{Acknowledgements.} 
This work was supported by the European Commission through the Marie Curie Initial Training Network PICQUE (Photonic Integrated Compound Quantum Encoding, grant agreement no. 608062, funding programme FP7-PEOPLE-2013-ITN, www.picque.eu) and through the project QUCHIP (Quantum Simulation on a Photonic Chip, grant agreement no. 641039, funding programme H2020-FETPROACT-2014, www.quchip.eu).

\textbf{Author contribution.} 
I.P., F.S. and G.C. fabricated and characterised the integrated optics devices. All authors discussed and contributed to the interpretation of the experimental results. I.P., G.C., A.C. and R.O. wrote the paper.

\textbf{Competing financial interests.}
The authors declare no competing financial interests.



\begin{thebibliography}{10}
\bibitem{NotaVetro} The glass used in this work is specifically EAGLE 2000, which is currently out of production, being substituted by the new version EAGLE XG. However, as far as the waveguide writing process by FLM is concerned, we have found no difference in the use of the two glasses. 
\end{thebibliography}
\end{document}